\documentclass[aoas,nameyear,dvips]{arximspdf}

\doi{10.1214/09-AOAS312D}
\referstodoi{10.1214/09-AOAS312}
\volume{3}
\issue{4}
\pubyear{2009}
\firstpage{1282}
\lastpage{1284}

\begin{document}
\begin{frontmatter}

\title{Discussion of: Brownian distance covariance}
\runtitle{Discussion}
\pdftitle{Discussion on Brownian distance covariance by G. J. Szekely and M. L. Rizzo}

\begin{aug}
\author[A]{\fnms{Andrey} \snm{Feuerverger}\ead[label=e1]{andrey@utstat.toronto.edu}\corref{}}
\runauthor{A. Feuerverger}
\affiliation{University of Toronto}
\address[A]{Department of Statistics\\
University of Toronto\\
180 St. George Street\\
Toronto, Ontario\\
Canada M55 363\\
\printead{e1}} 
\end{aug}




\end{frontmatter}

Concepts of dependence are central in the theory of statistics
and to most of its applications.
It is therefore a pleasure to commend the authors---hereafter
SR---for their theoretical contributions to our understanding
of some of its subtler aspects, and for their provocatively
interesting data analysis examples.

Standard measures, such as the product moment correlation,
Spearman's rank correlation, Kendall's tau
or Fisher--Yates' normal scores statistic, are all deficient.
These only measure dependence of a ``monotone character''
and will not be effective even in such simple situations
as when $Y$ has a nonmonotone regression on $X$ and $X$
is sampled randomly.
Another simple example where such measures fail is when
$X_i = V_i Z_i$ and $Y_i = V_i Z_i^\prime$,
where the ``innovations'' $Z_i$, $Z_i^\prime$ are, say, independent
standard  normal variates, but the $X_i$, $Y_i$ share a common
random scaling $V_i$; such structures arise in the stochastic
volatility models of finance.

Owing to their importance, consistent measures of dependence---and, in particular, measures which in principle admit sample
analogues on the basis of which tests consistent against all
dependence alternatives can be constructed---have appeared previously, and at least as far back
as Renyi (\citeyear{1953}).
Renyi's measure has, of course, ideal theoretical properties,
but implementing its sample analogues is not straightforward,
and for that reason it has not become a mainstay in applications.
[See, e.g., Buja (\citeyear{1990}).]
In that respect the dependence measure (which predate's Renyi's)
introduced by Hoeffding (\citeyear{1948}), and later rediscovered
in a more transparent form by Blum, Keifer and Rosenblatt (\citeyear{1961}),
has been more successful. See also Cs\"{o}rg\H{o} (\citeyear{1985}).

There is also some precedent for the measures proposed at (2.4)
and (2.6) in SR (at least for the case when $\alpha = 1$),
although these appear here in a substantially extended form,
and based on a novel approach with fresh interpretations.
For example, Feuerverger (\citeyear{1993})---hereafter F93---proposed measures based on
\begin{equation}\label{eq1}
\int \int
\frac{ |
f_{X,Y}^n(s,t) - f_X^n (s) f_Y^n (t)
 |^2  }
{(1-e^{-s^2})(1-e^{-t^2})}
 W(s,t) \,  ds \, dt,
\end{equation}
with $W(s,t)$ a suitable weight function.
In F93, the denominator term in (\ref{eq1}) was suggested on the basis
of its being (proportional to) the limiting variance
of the term within the modulus in the numerator,
under the null hypothesis of independence
in the case of standard normality.
The ratio within the integral in (\ref{eq1}) is defined by continuity
at the limiting values of $s=0$ and/or $t=0$.
Using the bell-shaped weight function
\begin{equation}\label{eq2}
W(s,t) = \biggl( \frac{1 - e^{-s^2}}{s^2}   \biggr)
 \biggl( \frac{1 - e^{-t^2}}{t^2} \biggr)
\end{equation}
leads to the form (2.6) of SR.
In fact, simple modifications to the weight function~(\ref{eq2})
can lead to interesting and potentially
useful variants of the $T_1, T_2, T_3$ statistics
defined by SR at (2.11),
wherein the absolute value functions are replaced
by more general functions;
see, for example, the computations leading up to (4.11) in F93.
It should be noted, however, that in F93 the variates $X$ and $Y$
are univariate, while SR deal with the case where these are
random vectors of dimensions $p$ and~$q$.

The case of univariate $X$ and $Y$ affords another advantage,
and, in particular, with respect to desiderata one might wish to place
on a dependence measure.
Thus, it is desirable that a dependence measure
should not require moment conditions on the variables---not even a finite first moment.
And second, it is desirable to go beyond the stated scale invariance
$(X,Y) \rightarrow (\varepsilon X, \varepsilon Y)$,
not only so as to allow the values of $\varepsilon$
applied to $X$ and $Y$ to differ, but also to have the invariance
$(X,Y) \rightarrow (\phi (X), \psi (Y))$ with respect to strictly
monotone transformations $\phi$ and $\psi$.
In the univariate case, this may be achieved by replacing
the $X$'s and $Y$'s by, say, their normal scores.
[In fact, this is the reason behind the choice of denominator in (\ref{eq1}).]
The resulting rank-type test will then also have the advantage
of being $H_0$-distribution free as SR note in Section 4.3.
Furthermore, since the empirical marginal distributions
will then no longer be random, the term $T_3$
in (2.11) of SR can then be dispensed with,
while the term $T_2$ can be reduced,
resulting in substantially simplified computations.
Of related note, the representation (2.8) of SR is
particularly interesting.

While it would certainly be of interest to examine what can be done
along such lines when $X$ and $Y$ are not restricted to be
univariate, there is a further problem of a multivariate character
that arises.
This refers to the case where we seek to assess mutual independence
among more than two variables, for example $X$, $Y$ and $Z$,
with each of these being either univariate or multivariate.
This problem was alluded to briefly in F93, and perhaps some
progress may be possible on the basis of decompositions along lines
indicated, for example, in Deheuvels (\citeyear{1981}).
However, this further multivariate problem still remains
largely unresolved.

Progress sometimes consists of seeing a familiar object
in a totally new way, and
the notion of Brownian covariance---as indeed of covariances
relative to other stochastic processes,
particularly the fractional Brownian motions---introduced in this
article by SR is particularly novel.
Certainly the fact that $\mathcal W$ should equal $\mathcal V$ seems
at least as surprising as SR purport it to be.
But no small part of that surprise stems from the fact
that Brownian covariance
should lead to a statistic that happens to be
consistent against all alternatives.
Why should this have happened?
What role does normality of the process play in it?
Could the essential condition for consistency be
that the process be of full rank in the sense of requiring
a complete set of basis functions to represent it?
Or is it enough that some separating class
of functions [e.g., Breiman (\citeyear{1968}), page 165ff]
should underly the process in some sense?
Here SR leave us with a nice mystery which seems
surely worthwhile to try to resolve.

\printaddresses


\begin{thebibliography}{99}

\bibitem[\protect\citeauthoryear{}{1961}]{1961}
\textsc{Blum, J. R., Keifer, J.} and \textsc{Rosenblatt, M.} (1961).
Distribution free tests of independence based on the sample
distribution function.
\textit{Ann. Math. Statist.} \textbf{32} 485--498.
\MR{0125690}

\bibitem[\protect\citeauthoryear{}{1968}]{1968}
\textsc{Breiman, L.} (1968). \textit{Probability}.
Addison-Wesley, Reading, MA.
\MR{0229267}

\bibitem[\protect\citeauthoryear{}{1990}]{1990}
\textsc{Buja, A.} (1990). Remarks on functional canonical variates,
alternating least squares methods, and ACE.
\textit{Ann. Statist.} \textbf{18} 1032--1069.
\MR{1062698}

\bibitem[\protect\citeauthoryear{}{1985}]{1985}
\textsc{Cs\"{o}rg\H{o}, S.} (1985). Testing independence by the empirical
characteristic function.
\textit{J. Multivariate Anal.} \textbf{16} 290--299.
\MR{0793494}

\bibitem[\protect\citeauthoryear{}{1981}]{1981}
\textsc{Deheuvels, P.} (1981). An asymptotic decomposition for
multivariate distribution-free tests of independence.
\textit{J. Multivariate Anal.} \textbf{11} 102--113.
\MR{0612295}

\bibitem[\protect\citeauthoryear{}{1993}]{1993}
\textsc{Feuerverger, A.} (1993). A consistent test for bivariate dependence.
\textit{Int. Statist. Rev.} \textbf{61} 419--433.

\bibitem[\protect\citeauthoryear{}{1948}]{1948}
\textsc{Hoeffding, W.} (1948). A non-parametric test of independence.
\textit{Ann. Math. Statist.} \textbf{19} 546--557.
\MR{0029139}

\bibitem[\protect\citeauthoryear{}{1953}]{1953}
Renyi, A. (1953). On measures of dependence. \textit{Acta Math. Hungar.} \textbf{10} 441--451.


\end{thebibliography}
\end{document}